\documentclass[3p,times]{elsarticle}

\usepackage{amssymb}

\usepackage[figuresright]{rotating}

\begin{document}



\author{Robert M. Ziff}
\ead{rziff@umich.edu}


\title{Results for a critical threshold, the correction-to-scaling exponent and susceptibility amplitude ratio for 2d percolation  \tnoteref{xx}}

 \tnotetext[xx]{Talk presented at 24th Annual Workshop, ``Recent Developments in Computer Simulational Studies in Condensed Matter Physics,"
Center for Computational Physics, University of Georgia, Athens, Georgia, Feb. 21-25, 2011}



\address{Center for the Study of Complex Systems and Department of Chemical Engineering, \\
University of Michigan, Ann Arbor, MI 48109-2136 USA}

\begin{abstract}
We summarize several decades of work in finding values for the percolation threshold $p_c$ for site percolation on the square lattice, the universal correction-to-scaling exponent $\Omega$, and the susceptibility amplitude ratio $C^+/C^-$, in two dimensions.  Recent studies have  yielded the precise values $p_c = 0.59274602(4)$, $\Omega = 72/91 \approx 0.791$, and $C^+/C^- = 161.5(2.0)$, resolving long-standing controversies about the last two quantities and verifying the widely used value $p_c = 0.592746$ for the first.
  \end{abstract}
\maketitle




\section{Introduction}
\label{Introduction}

Percolation concerns the formation of long-range connectivity in a system.  When the site or bond 
occupancy reaches a critical threshold $p_c$, infinite connectivity first forms.  For some systems (e.g., bond
percolation on the square lattice, or site percolation on the triangular lattice---all in two dimensions (2d)), $p_c$ is known exactly,
while for others, it must be determined numerically.  A notable example of the latter is site percolation on the square lattice, whose threshold has been estimated in dozens of studies, many quite extensive, since 1960.  In this paper we list some of the methods that have been used in making those estimates, and summarize the values that have been found.  We also summarize results for two related quantities---the correction-to-scaling exponents, and the susceptibility amplitude ratio, which have also been the subject of numerous studies over the last several decades.
\begin{table}[t]
\caption{Determinations of $p_c$ for site percolation on the square lattice. \ 
Numbers in parentheses
represent errors in last digit(s).}
\begin{center}
\begin{tabular}{|l|l|l|l|l|l|}
\hline
year & author & method & $p_c$\cr 
\hline
1960 & Elliott, Heap, Morgan  \& Rushbrooke  \cite{ElliottHeapMorganRushbrooke60}  & series& $0.48$  \cr
1961 & Domb \& Sykes  \cite{DombSykes61}  & series& $0.55$  \cr
1961 & Frisch, Sonnenblick, Vyssotsky, Hammersley  \cite{FrischSonnenblickVyssotskyHammersley61}  & MC& $0.581(15)$  \cr
1963 & Dean  \cite{Dean63}  & MC& $0.580(18)$  \cr
1964 & Sykes \& Essam  \cite{SykesEssam64}  & series& $0.59(1)$  \cr
1967 & Dean \& Bird  \cite{DeanBird67}  & MC& $0.591(5)$  \cr
1972 & Neal  \cite{Neal72}  & MC & $0.593(5)$  \cr
1976 & Sykes, Gaunt \& Glen  \cite{SykesGauntGlen76}  & series& $0.593(2)$  \cr
1976 & Stauffer \cite{Stauffer76}  & series& $0.591(1)$  \cr
1976 & Leath \cite{Leath76} & MC & $0.587(14)$ \cr
1978 & Hoshen, Kopelman \& Monberg \cite{HoshenKopelmanMonberg78}  & MC & $0.5927(3)$  \cr
1980 & Reynolds, Stanley \& Klein  \cite{HoshenKopelmanMonberg78}  & MC & $0.5931(6)$  \cr
1982 & Derrida \& de Seze  \cite{DerridadeSeze82}  & TM & $0.5927(2)$  \cr
1982 & Djordjevic, Stanley \& Margolina \cite{DjordjevicStanleyMargolina82}  & series& $0.5923(7)$  \cr
1984 & Gebele \cite{Gebele84}  & MC& $0.59277(5)$  \cr
1985 & Rapaport \cite{Rapaport85}  & MC& $0.5927(1)$  \cr
1985 & Rosso, Gouyet \& Sapoval \cite{RossoGouyetSapoval85}  & MC & $0.59280(1)$  \cr
1985 & Derrida \& Stauffer \cite{DerridaStauffer85}  & TM & $0.59274(10)$  \cr
1986 & Ziff \cite{Ziff86} & MC & 	$0.59275(3)$  \cr
1986 & Kert\'esz \cite{Kertesz86} & TM & $0.59273(6)$ \cr
1986 & Ziff \& Sapoval \cite{ZiffSapoval86} & MC & $0.592745(2)$  \cr
1988 & Ziff \& Stell  \cite{ZiffStell88} & MC & $0.5927460(5)$  \cr
1989 & Yonezawa, Sakamoto \& Hori \cite{YonezawaSakamotoHori89} & MC & $0.5930(1)$  \cr
1992 & Ziff  \cite{Ziff92} &	MC & $0.5927460(5)$  \cr
2000	  & Newman \& Ziff \cite{NewmanZiff00}	 &MC& $0.59274621(13)$  \cr
2003 & de Oliveira, N\'obrega \& Stauffer \cite{deOliveiraNobregaStauffer03} & MC & $0.59274621(33)$  \cr
2005	 & Deng \& Bl\"ote \cite{DengBlote05} & MC & $0.5927465(4)$  \cr
2007 & M. J. Lee  \cite{Lee07}  & MC & $0.59274603(9)$  \cr
2008 & M. J. Lee \cite{Lee08} & MC & $0.59274598(4)$  \cr
2008 & Feng, Deng \& Bl\" ote \cite{FengDengBlote08}  & TM/MC & $0.59274605(3)$  \cr
\hline
\end{tabular}
\end{center}
\label{table:pc}
\end{table}%

\section{The determination of $p_c$ for site percolation on the square lattice}

Many methods have been developed to estimate percolation thresholds numerically.  Some of the more popular ones include:
\begin{itemize}
\item Series analysis methods, which involve finding exact statistics for smaller clusters, and were especially important in the earlier years of study \cite{GauntSykes76,AdlerMoshePrivman82}.
\item The transfer-matrix method \cite{DerridadeSeze82,DerridaStauffer85}, which involves finding exact solutions in finite-width strips, and has been recently revived to find very accurate values of $p_c$ \cite{FengDengBlote08}.
\item  Finding the value of $p$ where the crossing probability (open system) or wrapping probability (on a torus) equals a known amount or becomes independent of the size of the system \cite{YonezawaSakamotoHori89,Ziff92}.   When using the wrapping probability on a torus, the convergence of the estimate $p_c(L)$ is particularly fast: for a square system of size $L$, one has $p_c(L) - p_c \sim L^{-11/4}$ \cite{NewmanZiff01}.
\item Using various real-space renormalization-group theory ideas \cite{ReynoldsStanleyKlein80, BallesterosEtAl99}.
\item  Iterative searching for the value of $p$ where crossing first occurs.  The average of that value is used as an estimate for $p_c$ \cite{StaufferAharony94}.
\item Adding bonds to a system one at a time until crossing or wrapping  first occurs.  Here one must convolve with the binomial distribution to find the usual fixed-$p$ (``grand canonical") ensemble \cite{NewmanZiff00,ZiffNewman02,GouldTobochnikChristian06}.
\item Frontiers or hull-walks in a gradient \cite{RossoGouyetSapoval85,ZiffSapoval86,SudingZiff99,QuintanillaZiff07}, in which the estimate comes from simulations carried out in single runs for a given gradient.  Valid in 2d only.
\item Statistics of individual cluster growth with a maximum size cutoff, which eliminates boundary effects.  Valid in all dimensions, but requires separate runs at different values of $p$ \cite{LorenzZiff98,LorenzZiff98a}.
\end{itemize}

\begin{table}[h]
\caption{Determinations of $\Omega$, $\omega = D \Omega = (91/48) \Omega$, and $\Delta_1 = \Omega/\sigma = (91/36) \Omega$. \ 
Numbers in parentheses
represent errors in last digit(s).}
\begin{center}
\begin{tabular}{|l|l|l|l|l|l|l|l|}
\hline
year & author & method & $\Omega$ & $\omega$  & $\Delta_1$\cr 
\hline
1976 & Gaunt \& Sykes  \cite{GauntSykes76}  & series  & $0.75(5)$ & 1.42 & 	1.90  \cr
1978 & Houghton, Reeve \& Wallace \cite{HoughtonReeveWallace78} & field theory & 0.54--0.68 & 0.989--1.28 & 1.32--1.71\cr
1979 & Hoshen et al. \cite{HoshenStaufferBishopHarrisonQuinn79} & MC & 0.67(10) & 1.27 & 1.69 \cr
1980 & Pearson \cite{Pearson80} & conjecture & $64/91$$\approx$$0.703$ & 1.333& 1.778 \cr
1980 & Nakanishi \& Stanley \cite{NakanishiStanley80}& MC & $0.6 \le \Omega \le 1$ & & \cr
1982& Nienhuis \cite{Nienhuis82}  & field theory & $96/91$$\approx$$1.055$ &2& 2.667 \cr
1982 & Adler, Moshe \& Privman  \cite{AdlerMoshePrivman82} & series& $0.5$ &0.95& 1.26\cr
1983 & Adler, Moshe \& Privman  \cite{AdlerMoshePrivman83} & series& $0.66(7)$ & 1.25& 1.67\cr
&\qquad '' & series for $p < p_c$& $0.49$ & 0.93& 1.24\cr
 1983 & Aharony \& Fisher \cite{AharonyFisher83,Adler85} & analytic correction & $55/91$$\approx$$0.604$ & $55/48$$\approx$$1.15$ & $55/36$$\approx$$1.53$\cr
1984 & Margolina et al.\ \cite{MargolinaDjordjevicStaufferStanley83,MargolinaNakanishiStaufferStanley84} & MC & $0.64(8)$ & 1.21& 1.62\cr
&\qquad " & series &  0.8(1) &1.52 & 2.02 \cr
 1985 &Adler \cite{Adler85} & series & 0.63 & 1.19 & 1.59 \cr
1986 & Rapaport\ \cite{Rapaport86} & MC & 0.71--0.74 & & \cr
1998 & MacLeod \& Jan \cite{MacLeodJan98} & MC  & 0.65(5) & 1.23 & 1.64 \cr
1999 & Ziff \& Babalievski \cite{ZiffBabalievski99} & MC  & 0.77(2) & 1.46 & 1.95\cr
2001 & Tiggemann \cite{Tiggemann01} & MC  & 0.70(2) & 1.33 & 1.77 \cr
2003 & Aharony \& Asikainen \cite{AharonyAsikainen03,AsikainenAharonyMandelbrotRauschHovi03}& theory, based on \cite{denNijs83} &$72/91$ & $3/2$ & 2 \cr
2007 & Tiggemann \cite{Tiggemann07} & MC  & 0.73(2) & 1.38 & 1.85 \cr
2008 & Kammerer, H\"ofling, Franosch \cite{KammererHoflingFranosch08}&MC  & 0.77(4) & 1.46 & 1.95\cr
2011 & Ziff \cite{Ziff11} & theory, based on \cite{Cardy06} &$72/91$$\approx$$0.791$ & $3/2$ & 2 \cr
\hline
\end{tabular}
\end{center}
\label{table:omega}
\end{table}%

In Table \ref{table:pc} we list the values that have been found for $p_c$.  On the average, the precision has improved about one digit per decade.  Each new digit generally requires about $1000$ times more work, both for statistical purposes (a factor of 100) and for quantifying finite-size corrections (about another factor of 10); this rate roughly corresponds to the rate of increase of computer power (speed and memory) over the years.  Two very precise recent works \cite{Lee08,FengDengBlote08} confirm that the value $p_c \approx 0.592746$, which was proposed more than 20 years ago \cite{ZiffStell88,Ziff92} and became a standard after its inclusion in \cite{StaufferAharony94}, is accurate to all significant figures---and in fact, the next digit is most likely a zero.  The average of the results of \cite{Lee08,FengDengBlote08} give $p_c = 0.59274602(4)$.

The square lattice site threshold is just one example; thresholds have been studied for scores of systems in various dimensions, and many of these results are summarized in the web-page:

\verb+http://en.wikipedia.org/wiki/Percolation_threshold+.

\noindent Examples of other systems that have been studied extensively include bond percolation on the kagom\'e lattice \cite{FengDengBlote08,YonezawaSakamotoHori89, vanderMarck97,ZiffSuding97,ZiffGu09,DingFuGuoWu10} and both site and bond percolation on 3d cubic lattices \cite{ZiffStell88,BallesterosEtAl99,LorenzZiff98,LorenzZiff98a,Grassberger92,AcharyyaStauffer98,JanStauffer98,SkvorNezbeda09,DammerHinrichsen04}.

\section{Corrections to scaling}
\begin{figure}[htbp] 
   \centering
   \includegraphics[width=4in]{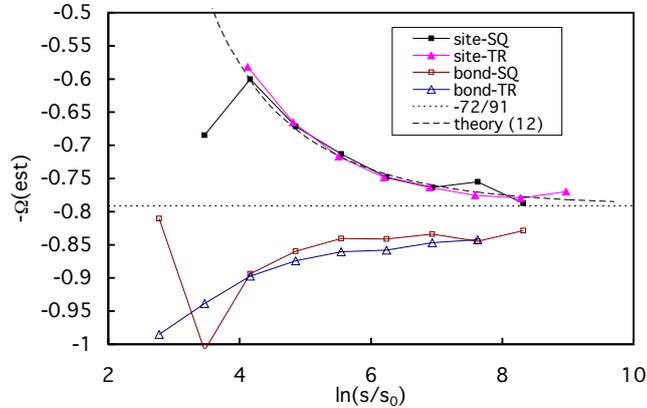} 
   \caption{Plot of $\Omega$(est) $\equiv -\log_2 [(C_s - C_{s/2})/(C_{s/2}-C_{s/4})]$
where $C_s = s^{\tau-2} P_{\ge s}$ and $P_{\ge s} = \sum_{s' \ge s} s' n_{s'}$ is the probability of
growing a cluster of size  $\ge s$.  Data from refs\ \cite{ZiffBabalievski99,Ziff11}. ``Theory" refers to Eq.\ (12) of \cite{Ziff11}.  
The non-universal metric factor $s_0$ equals  $0.25$ (site-square), $0.13$ (site-triangular), $0.25$ (bond-square) and $0.5$ (bond-triangular).
For large $s$, we see good evidence for the prediction $\Omega = 72/91$ \cite{AharonyAsikainen03,Ziff11}.}
\label{fig:omega}
\end{figure}

Many of the methods mentioned above depend upon knowing the behavior of the  corrections to the size distribution (number of clusters of size $s$) $n_s(p)$ at $p_c$, which are expected to be of the form:
\begin{equation}
n_s(p_c)  \sim A s^{-\tau} (1 + B s^{-\Omega}\ldots)
\end{equation}
where $\tau = 91/48$ is universal and is known exactly in 2d. $\Omega$ is also expected to be universal and has been studied in numerous works, usually in the context of determining another quantity such as $p_c$.   The resulting values are summarized in Table \ref{table:omega}.  The methods used include series analysis and Monte Carlo simulation on various systems, as well as predictions from theory.   There have been two recent conjectures that the value of $\Omega$ is exactly $72/91$, the first \cite{AharonyAsikainen03} based upon den Nijs' early result on the corrections to the correlation function \cite{denNijs83}, and the second \cite{Ziff11} based upon Cardy's more recent result for the crossing probability on an annulus \cite{Cardy06}.  Recent numerical results \cite{Ziff11} support  this value, as reproduced here in Fig.\ \ref{fig:omega}.  As seen in Table \ref{table:omega}, previous estimates and predictions have ranged from $0.49$ to $1.055$, but more recent measurements have been closer to the predicted $72/91 \approx 0.791$.  The exponents  $\Omega$,  $\omega = D \Omega$, and $\Delta_1 = \Omega/\sigma$ (where $D = 91/48$ and $\sigma = 36/91$) figure in many problems in percolation.

\begin{table}[t]
\caption{Determinations of the amplitude ratio $C^+/C^-$ for 2d percolation.}
\begin{center}
\begin{tabular}{|l|l|l|l|l|l|}
\hline
year & author & system, method & $C^+/C^-$\cr 
\hline
1976 & Sykes, Gaunt, Glen \cite{SykesGauntGlen76}  & lattice, series (12-20 order) & 1.3-2.0 \cr
1976 & Stauffer \cite{Stauffer76} & lattice, series analysis		 & 	$\approx 100$ \cr
1978 & Nakanishi \& Stanley \cite{NakanishiStanley78}	&	lattice, MC & 25(10) \cr
1978 & Wolff \& Stauffer \cite{WolffStauffer78} & lattice, series, fit $f(z)$ to Gaussian	 & 180(36) \cr
1979 & Hoshen, Stauffer, Bishop, Harrison, Quinn	\cite{HoshenStaufferBishopHarrisonQuinn79} & lattice, MC & 196(40) \cr
1980 & Nakanishi \& Stanley \cite{NakanishiStanley80}	&	lattice, MC (reanalyze) & 219(25) \cr
1980 & Aharony \cite{Aharony80}	&			$\epsilon = 6 - d$ expansion & 3.6Ð17 \cr
1981 &  Gawlinsky \& Stanley \cite{GawlinskyStanley81}  & 	overlapping disks, MC 	&	50(26) \cr	
1985 &  Rushton, Family \& Herrmann \cite{RushtonFamilyHerrmann85} & additive polymerization, MC& 	140(45) \cr	
1987 & Meir	\cite{Meir87} &			lattice, series & 210(10)\cr
1987 & Kim, Herrmann, Landau \cite{KimHerrmannLandau87}  & 	continuum model, MC	&14(10)\cr
1987 & Nakanishi	\cite{Nakanishi87}	 & 	AB percolation, MC		 & 		139(24) \cr
1988 & Balberg \cite{Balberg88}	 & 				widthless sticks, MC &		$\approx 3$ \cr
1988 & Ottavi	\cite{Ottavi87}		 &	approx.\ theory (Gaussian fit) & 193.9 \cr
1989 & Corsten, Jan \& Jerrard	\cite{CorstenJanJerrard89} & lattice, MC & 75(+40, -25) \cr
1990 & S. B. Lee \& Torquato	\cite{LeeTorquato90} & 		penetrable conc.\ shell &	1050(32)\cr
1990 & S. B. Lee 	\cite{Lee90} & 	disks, MC &	192(20) \cr
1991 & Hund \cite{Hund91}	 & 			random contour model, MC &	$\approx 200$ \cr	
1993 & Zhang \& De'Bell  \cite{ZhangDeBell93}	& Penrose quasi-lattice, series & 310(60) \cr
1995 & Conway \& Guttmann \cite{ConwayGuttmann95}	& lattice, series (26-33 order) & $45(+20, -10)$ \cr
1996 & S. B. Lee \cite{Lee96}		 & 			penetrable conc. shell, disks &	175(50) \cr	
1997 & S. B. Lee \& Jeon	\cite{LeeJeon97} & 		kinetic gelation, MC  & 	170(20) \cr	
1998 & Delfino \& Cardy \cite{DelfinoCardy98} 	 & 	theory, extrapolate Potts to $q=1$ & 74.2 \cr
2006 & Jensen \& Ziff \cite{JensenZiff06}	 & 	lattice, MC &  163(2) \cr
2006 & Jensen  \& Ziff \cite{JensenZiff06}	 & 	lattice, series & 162(3)\cr
2010 & Delfino, Viti \& Cardy \cite{DelfinoVitiCardy10}	 & 	theory, evaluate Potts at $q = 1$ & 160.2\cr
\hline
\end{tabular}
\end{center}
\label{table:amplituderatio}
\end{table}%
\section{Amplitude ratio}
Universality of the scaling function in percolation leads to the prediction that the ratio of the second-moment (or the ``susceptibility") for equal intervals below and above the threshold is a universal constant, written as $C^+/C^-$ (where ``below" is $+$ because of the mapping to the Potts model and corresponds to being above the transition temperature), also written as $\Gamma/\Gamma'$ and $\Gamma^-/\Gamma^+$.  Finding this ratio has been a notoriously difficult problem in percolation \cite{PrivmanHohenbergAharony91}, as can be seen in the summary in Table \ref{table:amplituderatio}.   Recently, three fundamental approaches---series analysis \cite{JensenZiff06}, Monte Carlo \cite{JensenZiff06}, and theory \cite{DelfinoVitiCardy10} have all converged to a remarkably consistent value of about $161.5(2.0)$.   Note that in some earlier work it was speculated that models of continuum percolation may have a different value for this ratio \cite{GawlinskyStanley81,Balberg88,LeeTorquato90}, but more recent work gave results consistent with this value \cite{Lee96}, confirming the expected universality.  Likewise, it appears that some kinetic percolation-like models, such as kinetic gelation, also have a similar susceptibility amplitude ratio.

\section{Conclusions}
We have seen that it has taken decades of work to find consistent values of $\Omega$ and $C^+/C^-$ for 2d percolation, and to 
find a very precise value of $p_c$ for site percolation on the square lattice.  Many other thresholds in various dimensionalities
are known to relatively low accuracy, and $\Omega$ and $C^+/C^-$ are not known to high accuracy in three and higher
dimensions, leaving much work for future studies.  Other amplitude ratios, such as those relating to the correlation length,
have not been studied to high accuracy, and controversies remain \cite{PrivmanHohenbergAharony91}.   Another area for future work is precise measurements and characterization
of the scaling function in all dimensionalities, not just of the amplitude ratios that follow from it.





\section{Acknowledgments}
This work was supported in part by the National Science Foundation Grant No.\ DMS-0553487. %

\bibliographystyle{elsarticle-num}
\bibliography{georgiaarxivV2}







\end{document}